\documentstyle[aas2pp4]{article}

\received{1996 April 2}

\slugcomment{Send preprint requests to ``jcc@cfht.hawaii.edu''}

\lefthead{Cuillandre et al.}
\righthead{Wide-field CCD imaging at CFHT: the MOCAM example}

\begin{document}

\title{Wide-field CCD imaging at CFHT: the MOCAM example}

\author{J.-C. Cuillandre\altaffilmark{1}, Y. Mellier\altaffilmark{2}, 
J.-P. Dupin and P. Tilloles}
\affil{Laboratoire d'Astrophysique de Toulouse Observatoire Midi-Pyr\'en\'ees,
    14 Av. E. Belin, 31400 Toulouse, France}

\author{R. Murowinski, D. Crampton and R. Wooff}
\affil{Dominion Astrophysical Observatory, 5071 West Saanich road,
    Victoria, B.C., V8X4M6, Canada}

\and

\author{G. A. Luppino}
\affil{Institute for Astronomy, University of Hawaii, 2680 Woodlawn Drive,
    Honolulu, Hawaii 96822}

\altaffiltext{1}{Visiting astronomer, Canada-France-Hawaii Telescope,
operated jointly by NRC of Canada, CNRS of France and the University
of Hawaii (Electronic mail: jcc@cfht.hawaii.edu).}
\altaffiltext{2}{Institut d'Astrophysique de Paris, 98 bis boulevard Arago,
75014 Paris, FRANCE.}

\begin{abstract}
We describe a new 4096$\times$4096 pixel CCD mosaic camera (MOCAM)
available at the prime focus of the Canada-France-Hawaii Telescope
(CFHT).  The camera is a mosaic of four $2048\times 2048$ Loral
frontside-illuminated CCDs with 15\thinspace$\mu$m pixels, providing a
field of view of $14'\times14'$ at a scale of 0$''\!\!.$21
pixel$^{-1}$.  MOCAM is equipped with B, V, R and I filters and has
demonstrated image quality of 0$''\!\!.$5--0$''\!\!.$6 FWHM over the
entire field. MOCAM will also be used with the CFHT adaptive optic
bonnette and will provide a field of view of 90$''$ at a scale of
0$''\!\!.$02 pixel$^{-1}$.  MOCAM works within the CFHT Pegasus
software environment and observers familiar with this system require no
additional training to use this camera effectively.  The technical
details, the performance and the first images obtained on the telescope
with MOCAM are presented. In particular, we discuss some important
improvements with respect to the standard single-CCD FOCAM camera, such
as multi-output parallel readout and dynamic anti-blooming.  We also discuss
critical technical issues concerning future wide-field imaging
facilities at the CFHT prime focus in light of our experience with
MOCAM and our recent experience with the even larger UH
8192$\times$8192 pixel CCD mosaic camera.
\end{abstract}


\section{Introduction}

With the coming of the new generation of giant telescopes, most 4-m
class telescopes must identify observational programs where they will
be the most competitive. Obviously, the outstanding image quality of
the prime focus Canada-France-Hawaii Telescope over a one square degree
field makes this telescope well suited for scientific programs
requiring wide-field imaging with good spatial resolution. In fact, for
some time there has been a general consensus that wide-field imaging
should be developed at CFHT (Crampton 1992) and the recent work on weak
gravitational lensing (Fahlman et al. 1994, Bonnet et al. 1994, Bonnet
\& Mellier 1995), and on the detection of dark halos and dust in
edge-on nearby galaxies (Lequeux \& Gu\'elin 1996a, Cuillandre et al.
1996) have clearly demonstrated the unique capabilities of this
telescope for high quality imaging over large fields. Strong scientific
interests ranging from solar system studies to cosmology have been
emphasized: e.g., detection of faint comets and Kuiper belt
objects, photometry and astrometry in clusters and of stars in the
galactic disk, statistical analyses of faint galaxy populations
(galaxy-galaxy correlation function, galaxy-galaxy lensing, luminosity
function of faint galaxies, weak lensing by clusters, superclusters and
large scale structure), and multicolor photometry of faint quasars.

Wide-field imaging with subarcsecond seeing requires large CCDs capable
of simultaneously covering a large field of view and  adequately sampling the
stellar point spread function.  In addition to being physically large,
with large numbers of pixels, the required CCDs should be of scientific
quality, with high sensitivity, low readout noise, excellent charge
transfer properties and good cosmetic quality. Moreover, the arrays
must be mechanically flat and stable. From this point of view, very large
monolithic arrays are potentially attractive detectors. Such devices,
like the four output Loral 9K$\times$9K 8.75\thinspace$\mu$m pixel CCD
(Bredthauer 1995), already exist and may be a promising approach,
particularly with the use of new microelectronic technologies which
permit single output readout at very high speed and very low noise
level (e.g. Luppino et al. 1995). At the present time, however, there
exist no large monolithic devices (thinned or unthinned) capable of
covering fields as large as the CFHT prime focus (1$^{\rm o}$ field or
$\sim$$\phi$300\thinspace mm), and such devices are not expected to
appear in the near future.  Mosaicing the CCDs in a large focal plane
is then a natural strategy to cover the wide field offered by the
telescope. The detection of baryonic dark matter in galaxy halos was
the first program to benefit from large mosaics: the MACHO project
(Stubbs et al. 1993) uses a dual 4K$\times$4K camera, and  the EROS
project (Arnaud et al. 1994) uses a 3K$\times$1K mosaic. There are now
several developments of large CCD mosaics already completed or still
underway (see Luppino et al. 1994 for a recent review).  The University
of Hawaii 8K$\times$8K mosaic, in operation at CFHT and UH 2.2m since
Spring 1995, is the largest astronomical CCD camera built to date
(Metzger et al. 1996).  However, the frontside-illuminated CCDs limit
the sensitivity of these cameras, particularly in the blue, and the
thinning of large devices is still a technical challenge despite many
attempts using various approaches. But now, 2K$\times$4K thinned CCDs
with $15\thinspace\mu$m pixels are proposed by several builders.
Future projects for wide-field imaging like the 8K$\times$10K thinned
mosaic (CFH10K), and the further development of a 16K$\times$16K
thinned mosaic which will cover the whole field of the CFHT prime focus
(Vigroux 1995) will take advantages of these rapid developments.

The MOCAM project received strong support at the third CFHT users
meeting in 1992 (Crampton 1992). A consortium was set up involving the
Dominion Astrophysical Observatory (DAO), the Institut des Sciences de
l'Univers (INSU), the Observatoire de Toulouse (OMP) and the University
of Hawaii (UH), with additional help from CFHT.  The camera was
inspired by the early UH4K mosaic camera (Luppino \& Miller 1992) with
strong emphasis on low cost and rapid construction that would result in
an easy-to-use, wide-field CCD camera for the astronomical community at
CFHT.  It was agreed with the CFHT executive that MOCAM should be
delivered as a full CFHT instrument compatible with the CFHT Pegasus
software interface environment (Christian et al. 1989). MOCAM was
actually only the first step towards the development of larger mosaics
for CFHT. In this respect, in addition to the strong scientific cases
that motivated its development, MOCAM permitted the analysis of many of
the technical and observational issues of wide-field imaging at CFHT.

This paper presents the MOCAM mosaic camera and some important issues 
relevant for the future of wide-field imaging at CFHT. Section 2 gives a
complete overview of the technical design and the performance of MOCAM
in the CFHT environment. Some details about image quality,
blooming and data handling are also presented to provide practical 
information for CFHT users. Section 3 discusses some aspects of wide-field 
imaging from our experience with MOCAM and the UH8K. This section
addresses important issues such as field distortion, scattered light,
etc., which become important when imaging very large fields.
We also point out various improvements needed at the CFHT prime focus
for it to accommodate even larger arrays in the future.

\section{MOCAM: a 4K$\times$4K MOsaic CAMera}

Since the UH 4K$\times$4K camera (Luppino \& Miller 1992) was designed
for the CFHT prime focus, it was decided to simply duplicate the dewar
and camera head, as well as the Loral 2K$\times$2K two-edge buttable
CCDs initially used in this camera.  The standard CFHT GenIII CCD
acquisition system based on a San Diego State University (SDSU) controller
(Leach 1988) was adapted by the Toulouse group to a parallel
multi-output system, and the Pegasus software was upgraded to allow the
acquisition of images as large as 33 Mbytes with the multi-output
feature.  The mechanical interface between the dewar and the prime
focus bonnette which handles a five position filter wheel and a large
iris shutter was designed and built at DAO.  The whole camera was fully
tested and optimized in Toulouse before shipment to CFHT for the
successful November 1994 first light on the telescope.

\subsection{CCDs and Dewar}

The CCDs designed by J. Geary (Smithsonian Astronomical Observatory)
were produced by the Loral Fairchild Imaging Sensor company (Geary et
al. 1990). These thick 2048$\times$2048, 15\thinspace$\mu$m square
pixel, two-edge-buttable CCDs have two low-noise outputs with a single
serial register than can be split in two.  These CCDs are fabricated on
100\thinspace mm wafers and have been laid out on the wafer with mirror
symmetry, each wafer providing two left-hand CCDs and two right-hand
CCDs. The mosaic is assembled from four CCDs such that all columns are
oriented vertically, with the upper two devices having their serial
registers at the top edge and the lower two devices with serial
registers on their bottom edge.  The mosaic then, must be assembled
from two left-hand and two right-hand devices. Each Loral foundry run
produced twenty wafers and, after three attempts, one run delivered
several high quality CCDs with a yield of about 20\%. The CCDs have a
MPP boron implant under a parallel phase and allow high pixel capacity in
partial-inverted mode while keeping the dark current as low as in the
fully inverted mode. In MOCAM, a single output is used per CCD but, in
case of failure, switching from the corner output to the edge output
can be easily achieved; the possibility of using each alternative
output is provided for in the DSP code.

\placefigure{fig1}
\begin{figure}[ht]
\plotone{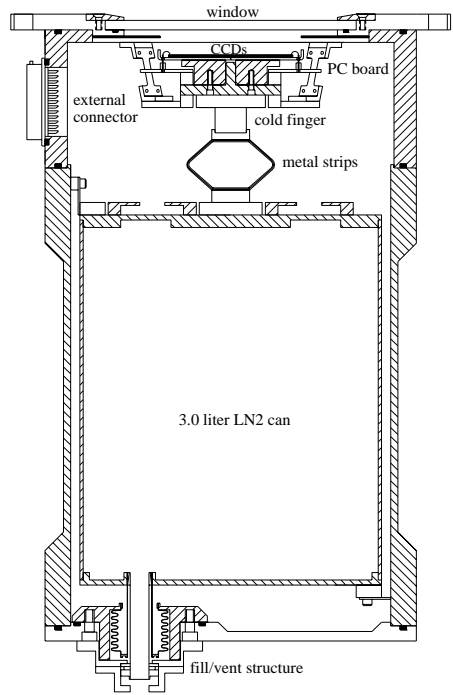}
\caption{Dewar assembly drawing.}
\end{figure}

The dewar design allows  four 2K$\times$2K two-edge buttable CCDs to be
mounted on a single alignment block upon which four kovar packages, to
which the CCDs are epoxy bonded, can be inserted and removed separately
(see Luppino \& Miller 1992).  It is then possible to test and optimize
each CCD individually before integration in the mosaic.  The four
packages are screwed on a cross-jig that secures the alignment of chips
to about $\pm$2 pixels and, most importantly, keeps the whole mosaic
flat to within $\pm30\thinspace\mu$m in accordance with the tolerance of the
depth of field at the CFHT prime focus ($60\thinspace\mu$m at $f$/4).
The gap between the CCDs is about 500\thinspace$\mu$m, or 30 pixels,
which only represents 0.8\% of the mosaic area. The mosaic size is
61\thinspace mm$\times$61\thinspace mm and offers a field of
$14'\times14'$ with a sampling of 0$''\!\!.$21 pixel$^{-1}$.

The nitrogen tank (Fig. 1) has a three liter capacity which keeps the 
focal plane at a temperature of $-135^{\circ}$C for more than 24 hours. 
A pressure of $10^{-5}$ Torr is achieved and
outgassing is extremely low, in part due to the absence of active 
electronics within the dewar. 

Optical ray tracing analysis shows that the effect on the image quality
of the curvature of the 5\thinspace mm thick, 140\thinspace mm diameter
quartz window due to the vacuum is negligible. However, the image
quality of the wide-field corrector does slightly degrade radially, so
it is preferable to focus towards the outer parts of the field to
insure the optimum image quality over the whole mosaic. 

\placefigure{fig2}
\begin{figure}[ht]
\plotone{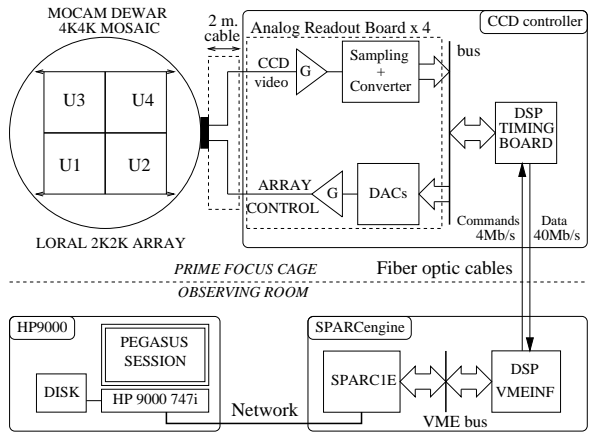}
\caption{MOCAM GenIII CCD controller architecture.}
\end{figure}

\subsection{CCD acquisition hardware architecture}

The CFHT GenerationIII controller (Kerr et al. 1994) is based on the
San Diego State University CCD controller and VME interface board
(Leach 1988). Figure 2 shows the system architecture where only one
analog card is represented.  A long shielded cable (2 m)
connects the dewar to the controller box which houses a utility board
and the four analog cards controlled in parallel by the timing board.
The temperature regulation and shutter timing are controlled
by the utility card.  Each analog card contains one video processor and
all digital to analog converters (DACs) to produce the various voltages
to run a CCD.  With such a configuration, each CCD of the mosaic is
completely isolated from the other chips, except for the ground plane,
hence strongly reducing the risks of interchip crosstalk and noise
injection.  Although the four CCDs are read out in parallel, the four
pixels are sent sequentially to the VME interface board through the
fiber optic on the host SUN SPARCengine 1E. The data are stored in a
shared memory segment and then descrambled and written to disk.

In the new CFHT release of the GenIII system, there is a dewar
identifier that prevents downloading and executing another chip DSP
code that could potentially damage the detector.

\subsection{The Pegasus CCD software}

The new CCD software takes advantage of the features of the last
Pegasus release and the use of a new powerful Hewlett Packard
workstation 9000-747i using the PA-RISC 7100 processor running at
100Mhz.  The Pegasus CCD software has been upgraded to allow fast and
efficient acquisition of large images from multi-output cameras.
Putting the data from all four chips into a single 33 Mbyte FITS file
is still reasonable and greatly simplifies the tasks of file managing,
archiving and preprocessing.  Also, it gives the observer the ability
to check the whole field in a single SAOimage window.  The drawback is
the extra time taken by the descrambling process that rearranges the
data into a single FITS file as they appear on the sky.  The latest
Pegasus release allows the data to be concurrently written to NFS disk
through the network as they are coming into the shared memory of the
SPARC1E.  Unfortunately it cannot descramble the data, and this process
must be implemented on the HP station.

\placefigure{fig3}
\begin{figure}[ht]
\plotone{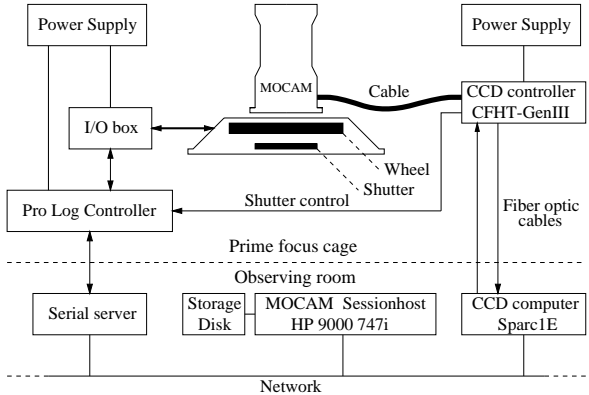}
\caption{MOCAM hardware.}
\end{figure}

There are four concurrent processes running in parallel:  $1)$ the DSP
program running on the timing board in the CCD controller that reads
out the mosaic in parallel and sends the data sequentially through the
fiber optic, $2)$ the DSP program running on the VME interface board in
the SPARC1E (the CCD computer) taking the data from the fiber optic and
putting them in the shared memory, $3)$ a Unix process running on the
SPARC1E, basically writing the data as fast as possible  from the shared
memory to the NFS disk (local disk for the HP) as the data fill the
shared memory, $4)$ a program on the HP that loads the incoming data into
memory data from its local disk and descrambles it. Once the last
pixel is read, all the data are already descrambled and the image
memory segment is then written to disk in less than 20\thinspace s.
The main limiting factor is the write-to-disk through the network.
The readout of the mosaic to the shared memory takes 170\thinspace s
(40\thinspace$\mu$s per burst of four pixels), whereas the NFS write to
disk is about 220\thinspace s long. The descrambler does not add time
and hence the complete readout time is 240\thinspace s.

This multi-output camera software is modular and can manage standard
single output systems or simply read out a single CCD within the mosaic
for raster selection and focusing.  This software can be extended to
larger mosaics like the future ten output CFH10K camera.  The Pegasus
user interface is identical to the standard FOCAM session and the
multi-output feature is invisible at this level.  The observers see
MOCAM as a monolithic 4K$\times$4K CCD where subraster selection is
possible as with any single detector,  except that subrasters that
span more than one CCD are not permitted. The interactive display of
the data is simplified as the observers have a full 4K$\times$4K image
in the SAOimage window at the end of each exposure. A complete
acquisition cycle within the Pegasus session, including the display
(20\thinspace s), takes less than 5 minutes.  Even with 33 Mbytes per
image, it is still convenient to save the data on DAT tapes or Exabyte
tapes which both provide storage capacities of at least 4 Gbytes.

As for all CFHT data, the images are archived at the Canadian Astronomy
Data Center (CADC).  The CFHT archiving system consists of a deamon on
a dedicated machine that saves, through the network, the data on a
optical disk.  Since all frames, regardless of their content, are
archived, a huge amount of data is accumulated during the MOCAM runs
and even with a 3 Gbytes per side capacity, the disk has to be changed
every four days.

\subsection{Filter wheel and shutter}

The filter wheel and associated electronics are based on a Pro-Log
controller running DAO's Universal Controller Kernel software, which
was designed, built and optimized at DAO. This software originates from
the HRCam project (McClure et al. 1989).  The filter position is
managed from the MOCAM Pegasus session through a serial server as shown
on Fig. 3.  This server gets the commands from the session host over
the network via the MUSIC client server protocol (Multi-User System for
Instrument Control developed at the Lick Observatory) handling the
Pegasus communications between all networked computers used at CFHT.

\placefigure{fig4}
\begin{figure}[ht]
\plotone{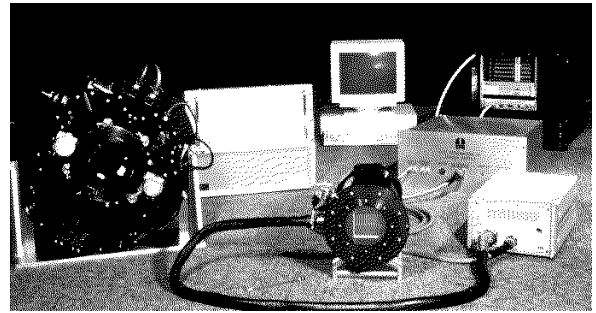}
\caption{Photograph of the whole assembled MOCAM camera.
The dewar is connected to the CCD controller with
its power supply box. At the rear, the VME crate containing the
SPARC1E and the HP9000 station holds the Pegasus software
for the data acquisition. On the left is the
filter wheel and its three associated boxes.}
\end{figure}

Two stepping motors control the filter wheel rotation and positioning
with a repeatability of 5\thinspace$\mu$m which is small compared to
the pixel size (15\thinspace$\mu$m). This guarantees a constant
flat-field pattern
which is easily removable by standard techniques. The time
required to select a new filter is about eleven seconds and a confirmation
message is sent to the user's session which allows the checking of the
instrument hardware status at any time.  The five positions of the
filter wheel presently hold four interference filters made by Barr
Associates: B, V, R and I; the transmission curves are shown in Fig. 5.
The filter size is 85\thinspace mm$\times$85\thinspace mm, much larger
than the mosaic size (61\thinspace mm), and the thickness ranges
from 4.6\thinspace mm for the V filter to 7.0\thinspace mm for the B
filter.

The shutter is under the control of the CCD controller to ensure
accurate timing. However exposures shorter than ten seconds suffer from
the slow mechanical response of the large iris shutter (diameter
150\thinspace mm) that opens in 50\thinspace ms and closes in
200\thinspace ms, the center part of the field being exposed
longer than the outside. Hence, very short exposures will yield
unreliable photometric data and are not recommended. 
Short exposures on bright standard stars should be avoided, but
defocusing these stars will lead to longer exposure times.
The standard dome and twilight flat-fielding techniques that use 
short exposure times become inaccurate for ultra-deep imaging, 
and superflats (Tyson 1990) are indispensable.  
Surma (1993) proposed a method to calibrate the shutter
response on short exposure images, but the superflat option remains
more efficient, providing better precision on both small scales and large
scales with a relatively small number of images.

\subsection{Optimization and performances}

\subsubsection{Hardware optimization}

Operating MOCAM in the prime focus cage of the CFHT is more difficult
than on the laboratory bench. The electronic environment of the
bonnette and the high-power electronics of the telescope and the dome
generate additional noise through electronic coupling. The main concern
was the long cable (2\thinspace m) connecting the CCD output to the
controller housing the pre-amplifier (Fig. 4).  This cable provides the
possibility of mounting the controller separately from the rotating
prime focus bonnette, but the controller was eventually mounted on the
side of the dewar for security and performance reasons.  During the
design of the camera, the challenging problem of controlling 
interference was
addressed, and a very highly shielded cable was built.  This cable
consists of four CCD harnesses completely insulated from each other to
prevent interchannel crosstalking through capacitive coupling.  Driving
the four CCDs completely separately with the use of independent analog
cards for each CCD avoids crosstalk between on-chip amplifiers which
can occur when the same voltage source is used for several outputs.

The ground plane was tested and adapted on the telescope.  The best
configuration is a ground spider centered on the backplane of the CCD
controller box while the dewar is turned into a Faraday cage by a thin
boron nitride layer between the nitrogen tank cold finger and the focal
plane. This layer ensures good thermal conduction while keeping the
CCDs electrically insulated from the dewar body.  All other power
sources are kept away from the CCD controller. A high-frequency
herringbone pattern appeared on bias frames at the telescope and were
found to be generated by interference between the readout and another
source. It was corrected by carefully adjusting the pixel frequency to
$\pm 3\thinspace\mu$s.  Eventually, with this highly-protected
environment, the noise in the bias frames was reduced to about 6
electrons, with neither small nor large scale patterns.

The detector security is a priority at this high altitude
(4200\thinspace m) where the dryness of the air often creates favorable
conditions for electrostatic discharge. The use of the dewar as a
Faraday cage was also motivated by this threat. Furthermore, 
voltage overshoots occurring at CCD controller power up and down are
particularly dangerous on this first version of the SDSU controller
used with MOCAM, as they inject voltages on the CCD well beyond the
rated tolerances of the components. Hopefully, the new generation of
SDSU controllers is designed to set up the voltages properly and safely.

The frequency of the thermal regulation cycle can generate additional
noise on the detector readout. We disabled the thermoelectric
regulation, preferring to adjust the number of metal strips of
various thickness that conduct the heat from the cold finger to the
nitrogen tank (Fig. 1). When used in the downward configuration at the
prime focus, the temperature stabilizes at $-135^{\circ}$C with
variations less than 2 degrees during the whole night, well below the
point where the temperature strongly influences the dark current
generation and quantum efficiency. The drawback of this method is the longer time required to
cool down the mosaic perfectly, almost 12 hours. The boron nitride
layer (0.5\thinspace mm thickness) has little influence on this
time.

\subsubsection{CCDs tests and performances}

The Loral foundry first produced calibration images at room temperature
for each chip when they were still on the wafer. This facilitated a
first selection of the best chip according to the apparent number of
bad cosmetics and CTE problems. The selected CCDs were cut and mounted
by Loral on kovar packages. Prior to integration in the mosaic focal
plane, each CCD was individually tested and optimized at DAO at
cryogenic temperatures.  Tests included gain and CTE measurements with a
$^{55}$Fe radioactive source, and linearity and cosmetic quality
evaluation with an integrating sphere. All these parameters were
optimized through the control of the parallel and serial transfer
clocking voltages, and the output amplifier bias and clocking
voltages.

All the tests had to be repeated at Toulouse once the best CCDs were put
together in the MOCAM dewar. In order to find the optimal behavior of
the four CCDs when run together, we considerably improved the
automation of the sequences, in particular on the optical bench, in the
data acquisition processes, and in the data reduction and data
analysis. This allowed us to secure a large number of measurements for
all sets of parameters automatically and to determine the optimum
performance of the camera with a high level of confidence.

Commonly used to determine the gain and the CTE, the $^{55}$Fe source
has the disadvantage of requiring a change of the quartz window with a
beryllium plate or a metal cover with the $^{55}$Fe source mounted
inside. To avoid changing the window, we also used a radioactive
tritium source (a twelve years decay $\beta$ radioactive source
interacting with a phosphorescent envelop generating optical photons)
with the quartz window to determine the gain, the linearity and the
full well pixel capacity through the photon transfer curve (Janesick et
al. 1987).  A large integrating sphere was used to test detector
uniformity, the parallel and serial CTE through the extended edge
response method, and to determine a map of cosmetic defects with low
level flat-fields.  The intercalibration of the four detectors showed
that the quantum efficiency is very uniform from one chip to another if
they originate from a same wafer lot. Measurements of the gain and CTE
with the tritium source appeared to be very close to the results obtained
with the $^{55}$Fe source and good enough for the precision required to
qualify a CCD for an astronomical use. The characteristics of the four
CCDs are very similar and summarized in Fig. 5.

As mentioned before, chip-to-chip crosstalk was a serious concern and
this topic was largely tested on the optical bench.  Various pickup
noises were canceled thanks to shielding and appropriate design of the
voltage plane. Simulating a very luminous star on one CCD allowed us to
set up properly the delays in the video processing sequence (Cuillandre
et al. 1995) and to eliminate crosstalk (ghost images) through
capacitive coupling to a level of 0.5\% between the video cables.
Aperture photometry on the ghost images indicates the crosstalk was
efficiently eliminated.

\placefigure{fig5}
\begin{figure}[ht]
\plotone{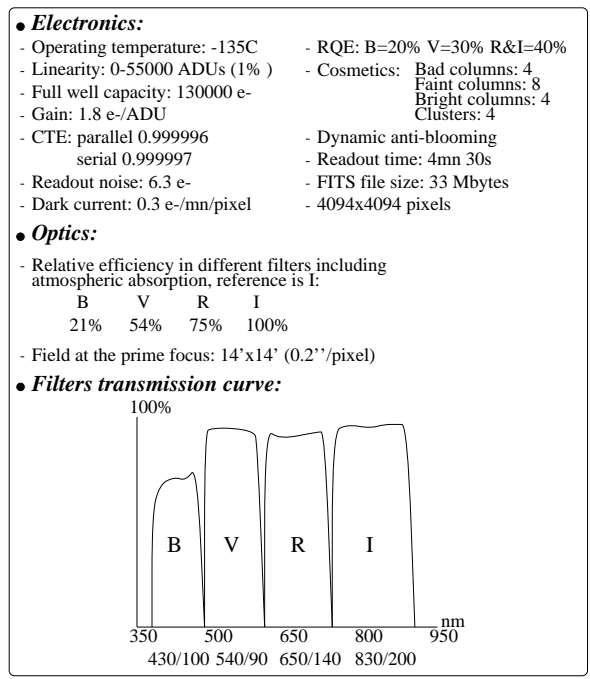}
\caption{Characteristics of the MOCAM CCDs and various camera
elements.}
\end{figure}

\subsubsection{Dynamic anti-blooming}

During the first observing run on the telescope in November 1994,
contamination by blooming from bright stars appeared to be a serious
limitation for the detection efficiency.  They actually contaminate a
large area of the CCD by charges spreading up and down along the
columns.  Janesick et al. (1992) proposed a new technique to cancel the
blooming by luminous stars, dynamic anti-blooming, that has been
successfully implemented by several observatories (e.g. Neely et al.
1993). Two parallel clock phases, P1 and P2 for our Loral CCDs, are
clocked during integration while the barrier phase P3, the doped
channel, remains inverted. This clocking sequence allows surplus
electrons to recombine with holes at the Si-SiO$_{2}$ interface when
the two collecting phases P1 and P2 invert.  The efficiency of this
technique, say the number of surplus charges that can be eliminated,
depends on the frequency of the clocking.

\placefigure{fig6}
\begin{figure}[t]
\plotone{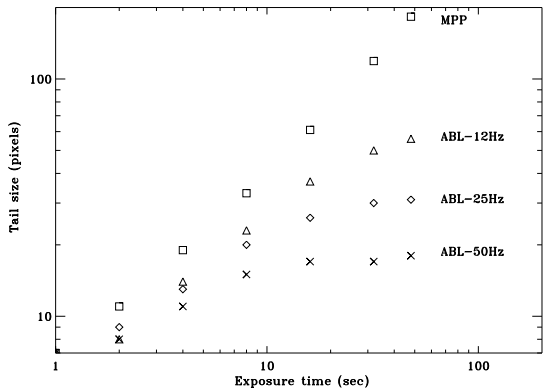}
\caption{Size of the blooming tail of a very luminous simulated star
versus exposure time for a set of anti-blooming frequencies.}
\end{figure}

Contamination by bright stars has two consequences, a halo and
blooming. The halo can be large and obliterates objects of scientific
interest within this area, while the blooming contaminates the data
much further away but only on a limited number of columns (Fig. 9,
left).  The optimum clocking frequency for MOCAM is chosen to reduce
the blooming within the halo of stars fainter than 8th magnitude on
exposures less than 30\thinspace mn long (the probability of finding an
8th magnitude star in a one degree field is close to one).  An
artificial star on an optical bench was used to measure the
effectiveness of dynamic anti-blooming as a function of frequency.
Figure 6 shows the size of the blooming tail versus exposure time for
different frequencies. The tail size increases  non-linearly as each
pixel acts as an individual pump, and the upper limit depends on the
star brightness and the frequency. The anti-blooming pumping efficiency
increases linearly over a wide range of anti-blooming frequencies
(12\thinspace Hz to 400\thinspace Hz) with a slope of 500
$e^{-}$s$^{-1}$pixel$^{-1}$Hz$^{-1}$, a typical value for Loral CCDs
(Kohley et al. 1995).  We found the 25\thinspace Hz frequency to be
optimal for our imaging application at the CFHT prime focus.  As
proposed by Janesick for the Loral CCDs (Janesick et al.  1992), the
optimal levels for the collecting phases are $-8\thinspace$V and
$+3\thinspace$V. This second level is higher ($+1.5\thinspace$V) than
the normal voltage used during charge transfer to improve the charge
collection.  The usual drawback of this technique is an increase of the
dark current and the generation of spurious noise due to the rising and
falling slopes of the two phases during the clocking. This noise is
uniform on the whole detector and increases linearly with the clocking
frequency with a slope of $2\times10^{-4}$
$e^{-}$s$^{-1}$pixel$^{-1}$Hz$^{-1}$.  This noise is remarkably low for
the MOCAM CCDs, but was about ten times higher for similar chips coming
from a previous run. The silicon quality seems a critical issue. CCDs
from the earlier run suffered from an abnormally high dark current when
operated at $-80^{\circ}$C.  Remarkably, dynamic anti-blooming decreased this
dark current by a factor of two over the standard MPP mode (see also
Janesick et al. 1992).

\placefigure{fig7}
\begin{figure}[t]
\plotone{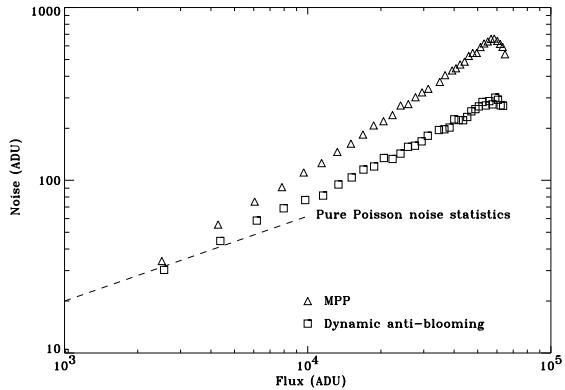}
\caption{Noise versus signal for the MPP mode and anti-blooming mode.}
\end{figure}

The behavior of MPP mode noise versus signal and anti-blooming 
mode (ABL) noise versus signal plotted on figure 7 is different.  The rough
equation for CCD noise (in ADU) at high flux is (Janesick et al. 1987):

\[ N = \sqrt { \frac{S}{g} + (e_{rqe} S)^{2} }\]

where $S$ is the average flux in the area in ADU, $g$ is the inverse gain in
$e^{-}$ ADU$^{-1}$ and $e_{rqe}$ is the quantum efficiency
non-uniformity from pixel to pixel.  This latter term is a result of
physical differences between the photosites of each pixel, and its
effect can be seen in figure 7 in the high signal regime as the data
departs from the line indicating pure photon statistics.  The value
for $e_{rqe}$ can be directly measured from the average of a large number
of uniformly-illuminated images, i.e., a flat-field. We found a typical value
of 0.9\% for the MPP mode and 0.4\% for the ABL mode. This difference
can be understood by examining the behavior of the photosites in the
two cases.  In MPP mode, the charge collection area is a result of the
barrier implanted under phase 3, and any lithographic or small scale
process variations during CCD fabrication in the definition of that
barrier will become manifest as $e_{rqe}$.  During normal (non-MPP) modes,
the charge collection area of a pixel is centered on and dominated by
the positive phase(s).  In ABL operation, which also has at least one phase
positive as in the normal mode, the center of the charge collection
area moves between two phases, which tends to
average out some of the process variations and reduce the value of
$e_{rqe}$.  It should be possible to further reduce this pixel response
non-uniformity by clocking three phases during an exposure, at the
expense of reduced resolution, as the average photosite during an
exposure is broadened.  In any case, the reduction of systematic errors
always leads to an improvement of the data quality.

\subsubsection{Tests and validation on the sky}

The first tests on the sky occurred in November 1994 during a fourteen
night run for Canadian and French weak lensing programs. Engineering
nights were also allocated to fully evaluate the instrument on the
sky.  Observations in the V, R and I filters of a large stellar field,
like SA 98 (Landolt 1992), showed that the image quality is uniform on
the whole mosaic. Observations of this field while keeping the
telescope motionless produced a large number of straight lines on the
mosaic which yielded a precise measurement of the misalignment of the
four chips. The deviation is about two pixels over the whole length of a chip
(0.03 degree).  The astrometry of the mosaic with respect to the
positions of the four quadrants in the single output FITS file was
easily obtained through two offset observations of the stellar field.

The intercalibration of the chips and the relative efficiency in
different filters were also made using the field of SA 98 (Fig. 5).  A
series of calibration images was also taken during the day using
dome-flats to build a photon transfer curve of the CCDs when mounted on
the telescope. This test was necessary to check that the response of
the detectors was still the same, despite substantial changes of the
environment, air temperature, humidity and noisy electronic neighbors.
A second run benefited from implementation of the dynamic anti-blooming
technique which proved to be very efficient on stars as bright as 8th
magnitude as shown on Fig. 9 (right) where the blooming is contained
within the halo area. The observation and the photometric analysis of
this stellar field (SA 110) both in MPP mode and anti-blooming mode
(only at the 25\thinspace Hz frequency) confirmed that the photometry
and the PSF are not affected.

\placefigure{fig8}
\begin{figure}[t]
\plotone{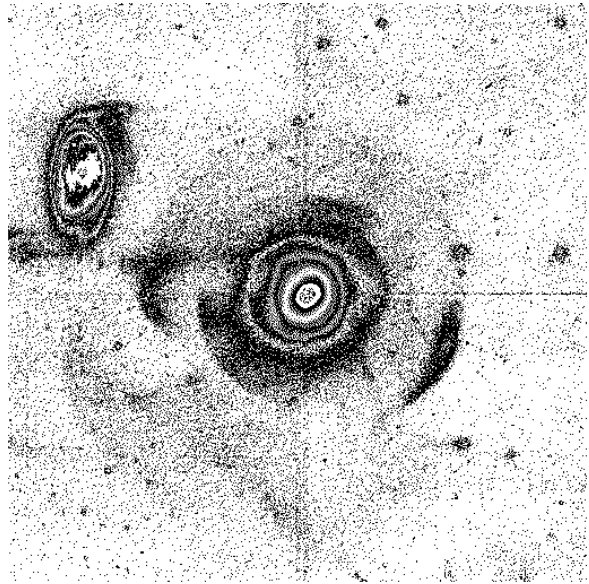}
\caption{``First light'' image of MOCAM (Nov. 94). The shell galaxy NGC 474,
the image quality is 0$''\!\!.$7 on the whole field of the V-band three hours
exposure (composition of six 30\thinspace mn exposures).
Outer shells magnitude approaches $\sim 27.5$ mag. arcsec.$^{-2}$.
A shift-and-add technique was used to fill the gap between the chips.}
\end{figure}

Thanks to the large amount of data obtained on blank fields, superflats
in different colors could be obtained.  Due to their specific file
format and the systematic loading of the whole image in memory,
standard software packages like IRAF or MIDAS are inadequate for
processing large images.  With the large size of each image and the
large number of images required to build a proper superflat, we
developed dedicated pre-processing FITS tools in C to allow fast and
easy pre-reduction of the data. These programs work with buffered
access to the file and do not require a lot of memory.

\section{Constraints of wide-field observations at CFHT prime focus}

A few months after the arrival and the first observations of MOCAM at
CFHT, the UH8K mosaic, a mosaic of eight 2K$\times$4K
frontside-illuminated Loral CCDs (Metzger et al. 1996), was used at the
prime focus of CFHT for solar system studies, weak lensing observations
and for the detection of dust in the outer regions of M 31 (Cuillandre
et al. 1996).  With a $28'\!\times\!28'$ field of view, this camera
covers 25\% of the prime focus field, offering a significant advantage
for these types of wide-field programs as compared to the field offered
by MOCAM.  Through the experience gained during the observations and
the reduction of the MOCAM and UH8K data, we found several constraints
and limitations due to the combination of the use of the prime focus,
its wide-field corrector and large CCD mosaics in general.

\subsection{Distortion and image quality}

The wide-field corrector has a well known radial distortion function
established from the analysis of photographic plates and reproduced
from raytrace analysis of the prime focus optics.  Based on the
notation used by Chiu (1976), the true radial distance $R$ (mm) can be
computed from the radial distance $r$ (mm) measured on the detector:
\[ R = r(1-9.04\times10^{-7}r^{2}-2.06\times10^{-12}r^{4})\] $r$ is
greater than $R$ and the corresponding true field angle on the sky in
arcminutes is $\theta=0.2288r$ where the multiplicative coefficient is
the scale at the center of the field in arcmin mm$^{-1}$.  This type
of distortion is common for wide-field correctors.  The presence of
this distortion eliminates the possibility of drift scanning with very
large mosaics with the present CFHT wide-field corrector.

\placefigure{fig9}
\begin{figure}[ht]
\plotone{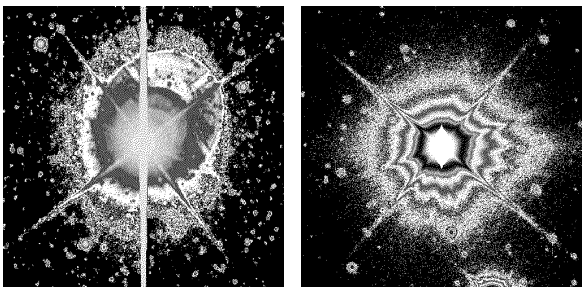}
\caption{Left: multiple reflections and blooming. Right: light diffusion and
anti-blooming on a 8.5th magnitude star in V.}
\end{figure}

The geometry of the field is an important issue for scientific programs
demanding accurate positions of objects.  Fortunately, this distortion
acts only at large scales and the local PSF is not affected.  This
point was demonstrated from the high image quality data obtained with
MOCAM and the UH8K. 
However the image quality is more sensitive in the
outer parts of the field, so focusing should be carried out near the
edges of the mosaic. Whatever the procedure used, focusing still
remains among the most crucial and delicate processes at CFHT. 

Strong distortion can also have an impact on the shift-and-add procedure. If
the offset is too large, the distortion will cause objects not to line
up perfectly when attempting to build a composite image with just
vertical and horizontal displacements.  One could avoid this problem by
keeping the shifts small (less than or of order 10$''$), but this will
result in poor superflats and in not filling the gaps between mosaic
elements. In practice, resampling the images is inevitable, to
correct both for the slight offsets and orientations of the individual
mosaic elements, and for the geometric distortion. This task
requires powerful computing facilities, and accurate astrometric
calibration of the mosaic.

\placefigure{fig10}
\begin{figure}[t]
\plotone{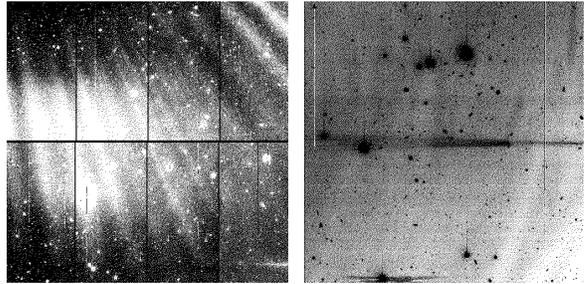}
\caption{Left: critical light diffusion in the CFHT wide-field corrector,
8K mosaic frame of the Cl 0024 cluster ($28'\!\times\!28'$ field). The 
4th magnitude star is 1$^{\circ}$ top left from the field.
Right: typical light reflection in the outer part of the field, the
star is about 5$'$ right from the edge of the mosaic. The horizontal
beam extends over 7$'$.}
\end{figure}

\subsection{Reflection and diffusion}

Another limitation of wide-field imaging on large telescopes comes from
off-axis reflections and scattered light from bright stars.  Over the
whole sky, the probability of finding an 8th magnitude star in a one
degree field is close to one. Brighter stars are often found in the
neighborhood of the target field, as we experienced during the MOCAM
and UH8K observations.  If the star is in the camera field of view, the
main consequence will be a large halo due to multiple reflections
between the filter, the dewar window, the CCD itself, and the optical
elements of the wide-field corrector (Fig. 9, left).  When the star is
outside the field, reflections on some mechanical parts of the
wide-field corrector and/or off-axis prime focus guider create beams
that seem to enter the dewar horizontally as shown in Fig. 10. These
observations demonstrate that the high dynamic range and sensitivity of CCD
detectors uncover factors that could limit the
performance of future observations. Diffuse light is presently a strong
limitation factor in deriving reliable flat-field calibration images
that allow flattening the data to better than $10^{-3}$ from the sky
background. Proper baffling of some part of the optical path as
proposed by Grundahl et al. (1996) may improve the correction to
about $10^{-4}$ which would be a major improvement of the detection
limit and would open new possibilities for the study of extended, faint
surface brightness, objects.

\section{Conclusion}

The CFHT now has a full wide-field imaging facility working in the
Pegasus environment that can be used by any observer.  Its performance
for astronomy is basically the same as the usual FOCAM instrument
equipped with thick CCDs, but with further attractive upgrades: a field
of view which is four times larger and dynamic anti-blooming,
multi-output, fast readout features.  The first observations at prime
focus have shown that the image quality is excellent over the whole
field and in all the wavelength range (from V to I). Images as good as
FWHM=0$''\!\!.$42 have been achieved by Fahlman et al. (1995, private
communication).  The MOCAM FITS files are 33 Mbytes each, but do not
require special computing facilities, nor dedicated software. Although
we emphasised that buffered access to the files provide more tractable
access to data, standard software like IRAF or MIDAS can also be used
if the mosaic is cut into four images.  First scientific results on
faint halos around edge-on spiral galaxies (Lequeux et al. 1996b)
demonstrate that MOCAM is a competitive instrument.  Its future use on
the CFHT adaptive optic bonnette (Arsenault et al. 1995) at the
cassegrain focus will provide a field of 90$''$ at a scale of
0$''\!\!.$02 pixel$^{-1}$.

Our experience gained in wide-field imaging at CFHT with MOCAM and UH8K
help to improve the imaging facility at CFHT, and  to prepare the
arrival of the next generation of large CCD mosaics.  In the near
future, the Pegasus system will incorporate a real time core that will
allow faster data acquisition. The network will no longer be used and
the data will go directly from the CCD controller to the HP memory.  In
the longer term, improvements of the wide-field corrector and the
telescope baffling are indispensable to reduce scattered and diffuse
light in the telescope.  This will considerably lower the background
light and the flat-field residuals. The guide probe appeared to be a
real problem in fields where a guide star couldn't be found outside the
camera field. The resulting vignetting on the mosaic obliterates the
data. Further developments for star selection based on deeper guide
star catalogs will be required, or even the design of a new bonnette,
particularly with the arrival of the one degree field 16K$\times$16K
camera.  If these improvements are made, it will be possible to make
ultra-deep wide-fields surveys at the CFHT prime focus.

The authors wish to thank Bernard Fort for his essential initial
impulse to the project, the Loral Fairchild Imaging Sensor company for
their support to our project, and all the CFHT staff, particularly
Steve Massey for his help during the setups on the telescope and Tim
Abbott for his contribution to the commissioning of MOCAM. We wish to
thank all the members of the instrument groups who have contributed to
this project:  P. Couderc, G. Delaigue and H. Valentin at OMP, B.
Leckie and A. Moore at DAO and R. Arsenault, B. Grundseth, D. McKenna,
S. Milner and J. Thomas at CFHT.

\end{document}